\documentclass[10pt,letterpaper]{article}
\usepackage{opex3}
\usepackage{bm}
\newcommand{\by}{$\times$}
\newcommand{\um}{${\bf \mu}$m\ }

\begin{document}

\title{Photonic Crystal Spectrometer}

\author{Nadia K. Pervez,$^{1,*}$ Warren Cheng,$^1$ Zhang Jia,$^1$ Marshall P. Cox,$^1$ Hassan M. Edrees,$^{1,2}$ and Ioannis Kymissis$^1$}

\address{$^1$Columbia Laboratory for Unconventional Electronics, Department of Electrical Engineering, \\
Columbia University, New York, 10027, USA}

\address{$^2$Department of Electrical and Computer Engineering, \\
The Cooper Union for the Advancement of Science and Art, New York, 10003, USA}

\email{$^*$nadia.pervez@gmail.com} 



\begin{abstract*}
We demonstrate a new kind of optical spectrometer employing
photonic crystal patterns to outcouple waveguided light
from a transparent substrate. This spectrometer consists of an
array of photonic crystal patterns, nanofabricated in a polymer on a
glass substrate, combined with a camera. The camera
captures an image of the light outcoupled from the patterned
substrate; the array of patterns produces a spatially resolved map of
intensities for different wavelength bands. The intensity map of the
image is converted into a spectrum using the photonic crystal pattern response
functions. We present a proof of concept by characterizing a white LED
with our photonic crystal spectrometer.
\end{abstract*}

\bigskip


\section{Introduction}

The invention of compact, low cost spectrometer modules has broadened the traditional spectrometry application base well beyond basic scientific measurement to include consumer applications such as monitoring and feedback in color printing and color paint matching \cite{productfocus,paint}. In a conventional spectrometer the relationship between diffraction angle and wavelength couples module size and resolution. Here, we present a new kind of spectrometer which  breaks that paradigm. The photonic crystal spectrometer measures a spectrum through the projection on to a basis formed by  photonic crystal pattern response functions rather than binning spatially resolved diffracted light. We present a proof of concept of this photonic crystal spectrometer by characterizing a white LED using an array of nine photonic crystal patterns.

A number of methods for making compact spectrometers have been previously proposed, including using the change in absorption of a material medium as a function of wavelength \cite{fluorescencedetection}, and monolithic integration of gratings on top of \cite{thinplanar} or at the end of \cite{sion-slab} thin planar waveguides. Another kind of spectrometer which also takes advantage of photonic crystals has also been reported \cite{superprism}; that device employs the enhanced dispersion of the superprism effect to create a spatially resolved pattern that can be read by a single detector. The superprism effect-based spectrometer is well suited to narrow bandwidth, high resolution applications. A two dimensional array based spectroscopy scheme using dielectric microsphere resonators has also been reported \cite{microresonator}; that array has a similar geometric scaling as the device presented here. Three important differences between this work and \cite{microresonator} are deterministic control over the array response, size -- the entire array presented here is smaller than a single microresonator array element -- and the calculation of a broadband spectrum based rather than use of the array to match single wavelength fingerprints. The work presented here demonstrates an extremely simple, inexpensive, and reproducible broadband spectrometer based upon the ability of photonic crystals to wavelength-selectively outcouple waveguided light from a substrate.

\begin{figure}[htbp]
\centering\includegraphics[width=10cm]{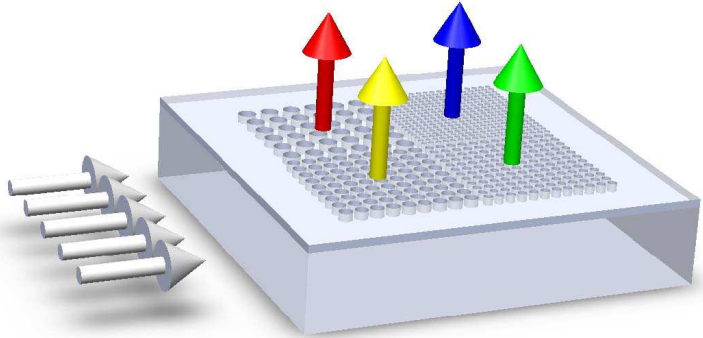}
\caption{Illustration of spatially resolved, wavelength-selective outcoupling of waveguided light using and array of photonic crystals. Light waveguided into the edge of the substrate is outcoupled by photonic crystal patterns corresponding to different bands of wavelengths.}
\end{figure}

Figure 1 illustrates the concept of spatially resolved, wavelength-selective outcoupling used in the spectrometer. In the illustration, light is introduced into the edge of the transparent substrate. Photonic crystals with four different lattice constants are patterned in the layer above the substrate. Each of these patterns outcouples a different band of wavelengths, illustrated by the arrows with different colors. A camera takes a picture of the substrate, which is used as an intensity map to quantitatively assess the response of the different photonic crystal patterns.

\section{Photonic crystals for outcoupling of waveguided light}

In a photonic crystal, a periodic potential formed by spatial variation in the relative permittivity $\epsilon_{r}$ of a medium interacts with electromagnetic radiation resulting in partial or complete bandgaps \cite{photonicxtal,squarelatticebandgap}. The band structure is determined by the choice of lattice, the basis formed by the shape and size of the holes, the thickness of the patterned layer, and the contrast in the spatial variation in $\epsilon_{r}$. The energy scale for the band structure is set by the lattice constant $a$. 

Similar two dimensional photonic crystal structures have been extensively investigated as a way to improve light extraction efficiency in light emitting diodes. Schemes involving the outcoupling of waveguided light using weak photonic crystals as well as the disruption of waveguided modes in the emissive layer using strong photonic crystals have been reported in the literature \cite{gapped-extraction, leaky-extraction1, leaky-extraction2, extraction-review}. The extraction scheme employed here is an extension of the waveguided light outcoupling scheme used for LED enhancement.

When light of wavelength $\lambda$ is introduced into a dielectric slab, the light has $|\bm{k}| = nk_{0}$ where $k_{0} = \frac{2\pi}{\lambda}$ and $n$ is the slab's index of refraction. In the slab, the magnitude of the transverse component of $\bm{k}$ is $|\bm{k_{\| m}}| = \sqrt{(nk_{0})^{2}-k_{\bot m}^{2}}$ where $k_{\bot m} \approx \frac{m \pi}{d}$, $d$ is the slab thickness and $m=1,2,3\ldots$. In a photonic crystal patterned region with lattice constant $a$ the periodic potential opens up gaps, allowing the transverse component of the wavevector to be scattered by a reciprocal lattice vector $\bm{G}$. Figure 2 illustrates the effect of this scattering, which allows previously guided modes to be extracted after a scattering event. The modes scattered to $|\bm{k_{\| m}}| > nk_{0}$ are evanescent in both the slab and air. The modes scattered to $k_{0} < |\bm{k_{\| m}}| < nk_{0}$ remain guided in the slab, and those scattered to $|\bm{k_{\| m}}| < k_{0}$ can propagate in the air. An additional constraint in our system is the ability of the camera to capture the light, limited to $|\bm{k_{\|m}}|< NAk_{0}$ where $NA$ is the numerical aperture of the camera objective. Only light that is imaged by the camera contributes to the measured photonic crystal response function. Thus the $NA$ affects the response function profile, with larger $NA$ objectives yielding wider response functions. The most efficient extraction occurs when $|\bm{k_{\|}}| \approx |\bm{G}|$. In our device an 0.96\thinspace mm thick glass coverslip is used as a substrate; for all but very high modes in our system $nk_{0} \gg \frac{m \pi} {d}$. Therefore $|\bm{k_{\|}}| \approx nk_{0}$, and optimum extraction occurs for $nk_{0} \approx |\bm{G}|$ or $\frac {\lambda}{n} \approx a$.

\begin{figure}[htbp]
\centering\includegraphics[width=10cm]{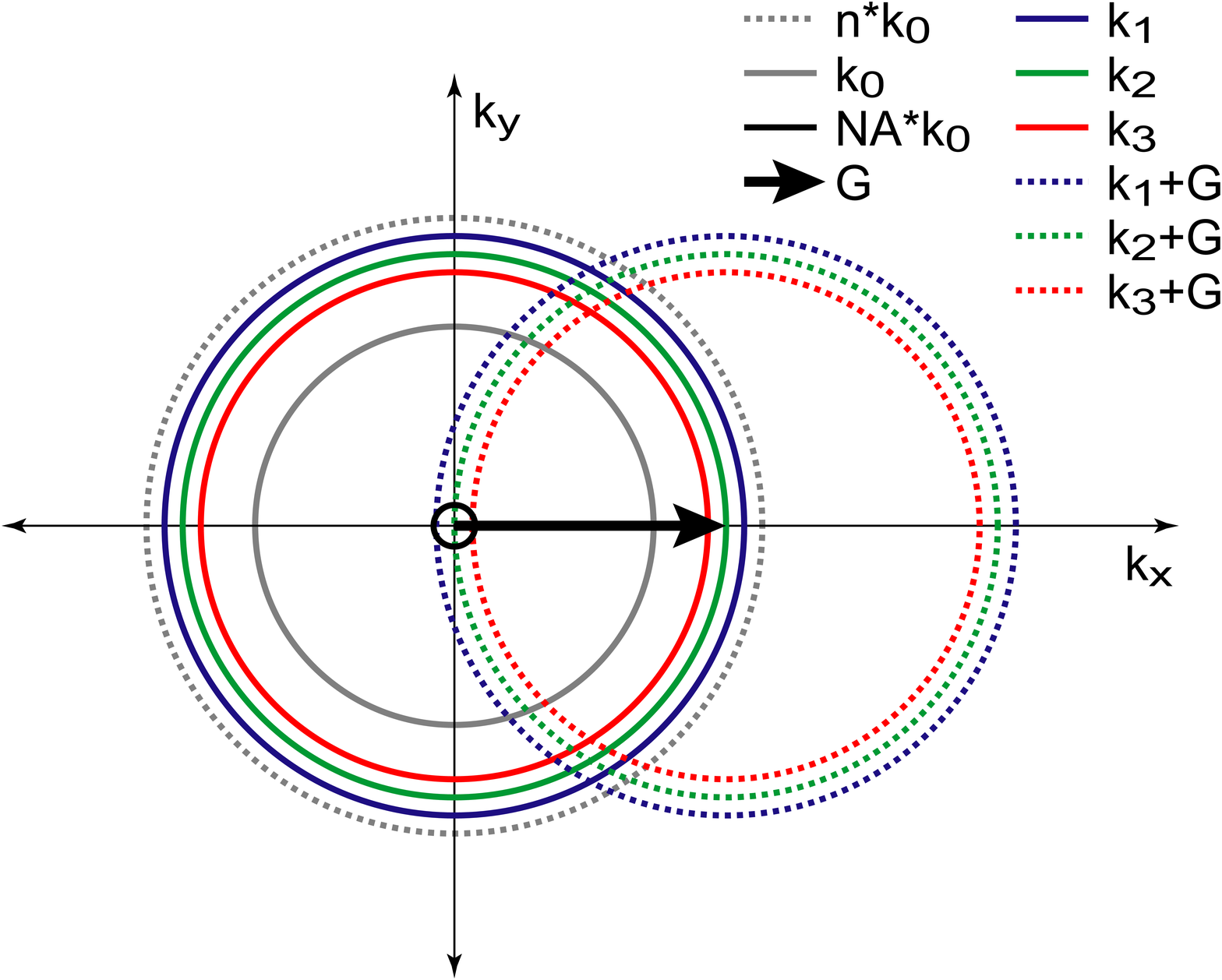}
\caption{Scattering of three guided modes with $|\bm{k_{\|}}| = \sqrt{k_{x}^{2} + k_{y}^{2}}=k_{1}$ (solid blue), $k_{2}$ (solid green), and $k_{3}$ (solid red) by a reciprocal lattice vector $\bm{G}$ into $|\bm{k_{1}}+\bm{G}|$ (dashed blue), $|\bm{k_{2}}+\bm{G}|$ (dashed green), and $|\bm{k_{3}}+\bm{G}|$ (dashed red). Modes scattered (i) outside of the dashed gray circle ($|\bm{k_{\|}}|>nk_{0}$) are evanescent in the slab, (ii) into the annulus bounded by the dashed gray and solid gray circles ($nk_{0}>|\bm{k_{\|}}|>k_{0}$) are guided in the slab and evanescent in air, (iii) into the annulus bounded by the solid gray and black circles ($k_{0}>|\bm{k_{\|}}|>NAk_{0}$) propagate in air but cannot be imaged by the camera, and (iv) inside the solid black circle ($|\bm{k_{\|}}|<NAk_{0}$) propagate in the air and can be imaged by the camera.}
\end{figure}

In the device presented here, the photonic crystals are fabricated in poly methyl methacrylate (PMMA) on top of the glass substrate. The index of refraction $n$ of PMMA over the wavelengths of interest (450 -- 700\thinspace nm) is 1.499 -- 1.512 which is slightly lower than the glass ($n \approx$ 1.52); the contrasting medium is air. The photonic crystal patterns employed are all square lattices, with fill factors ($r/a$, where $r$ is the hole radius) of approximately 0.34. With these design parameters, the periodic potential should not be strong enough to open a complete bandgap \cite{squarelatticebandgap}. The wavelength selectivity of the outcoupling is robust to disorder in the periodic potential since the interaction is distributed over several periods. SEM images of patterns similar to the smaller ones used in this work (data not shown) show non-uniform hole size as well as missing holes due to dose fluctuations in the electron beam lithography tool. 

\section{Experimental}

Arrays of photonic crystal patterns were fabricated in PMMA (MicroChem 495PMMA A series) on glass via electron beam lithography (EBL). The glass substrates were 22\thinspace mm\by 22\thinspace mm\by 0.96\thinspace mm coverslips. In this work square lattices photonic crystal patterns were chosen over hexagonal lattices for ease of fabrication. All of the lattice points written as large dots during EBL. An 8--10\thinspace nm layer of gold was used to prevent charging during the EBL process. A typical pattern exposure dose was 320\thinspace $\mu$C/cm$^{2}$.

To determine the optimal parameters for the photonic crystal array elements different lattice constants, fill factors, and PMMA thicknesses were evaluated in the context of light extraction measurements. Lattice constants from 250 to 650\thinspace nm were investigated to confirm response function peaks at $a \approx \frac {\lambda}{n}$. Fill factors ranging from 0.25 to 0.5 were assessed in the context of response function peak intensity and uniformity over the patterned regions (particularly in the context of proximity effects in EBL dose arrays). PMMA thicknesses from 233 to 509\thinspace nm were also evaluated for response function peak intensity but this process did not identify a single optimum value for all lattice constants. Based on these experiments, an array of nine square lattices with lattice constants from 300 to 420\thinspace nm and approximate fill factors of 0.34 were fabricated in 396\thinspace nm PMMA. The PMMA thickness was measured via spectroscopic ellipsometry and fill factors were estimated from SEM images. Each spectrometer pattern consisted of a 3\by 3 array of 30\um\by 30\um photonic crystal patterns, with a  total array size of approximately 100\um\by 100\um.

Photonic crystal array measurements were performed using waveguided light and a USB camera. Light was waveguided into all four edges of a sample using a custom fiber optic line of light assembly (Lumitex Inc. OptiLine). The line of light assembly consisted of a bundle of 310 0.010'' plastic fibers affixed into four 0.80'' linear array segments. Each linear array segment was butt-coupled to an edge of the glass substrate. Mineral oil was used to improve coupling between the line of light assembly and the glass. Although the line of light assembly was used for all of the array measurements in this work, we have coupled to similar structures using a single 1\thinspace mm plastic fiber. The line of light was required for the measurements presented here because the sample contained an EBL dose array of spectrometer patterns; the spectrometer pattern used in this work was therefore one of many fabricated on the same substrate. The line of light assembly allowed for simultaneous coupling into all patterns on the substrate. 

RGB images of the array were taken using a USB camera (Matrix Vision mvBlueFox-120c) using a 4\by objective with a numerical aperture of 0.096 (Infinity Photo-Optical Infinistix). As discussed in the previous section, the numerical aperture plays an important role in determining the response function widths. The pattern spectral responses were characterized using a programmable light source (Newport Apex Xe arc lamp and monochromator). The programmable light source (PLS) output was measured using a calibrated photodetector (Newport UV-818). The USB camera was characterized by taking images of the PLS output reflected from a teflon block. The teflon block reflectivity was characterized using the PLS and the calibrated photodetector. 

The performance of the photonic crystal spectrometer was characterized by replacing the PLS input with a 5\thinspace mm white LED driven at 5\thinspace mA. The same white LED at the same drive current was used for comparison measurement with a commercial spectrometer (Ocean Optics USB4000). The tip of the LED package was placed against SMA input for the comparison measurement. The response of the commercial spectrometer was characterized using the PLS. All measurements were performed in the dark. Image analysis was performed using Labview Vision (National Instruments). Nine 8\by 8 pixel regions of interest (ROIs) were defined for the 3\by 3 spectrometer array. The neutral weight grayscale luminance, rather than the green weighted NTSC grayscale luminance, is calculated for each ROI by averaging the mean red, green, and blue values in each region.  

\section{Results and Discussion}

Figure 3 shows the response of the array of nine photonic crystal patterns to four different light inputs. Figure 3(a)--(c) are narrow-band inputs; the multiple illuminated regions in each of these pictures illustrate the overlap between the response functions of the photonic crystal patterns. The full set of images is used to calculate the neutral weight grayscale luminance for each pattern. In this manner the response of each pattern to each measured wavelength is obtained, as shown in Figure 4. While the implementation presented here uses color images to obtain pattern intensity information, the same information can be readily obtained through the analysis of grayscale images.

\begin{figure}[htbp]
\centering\includegraphics[width=7cm]{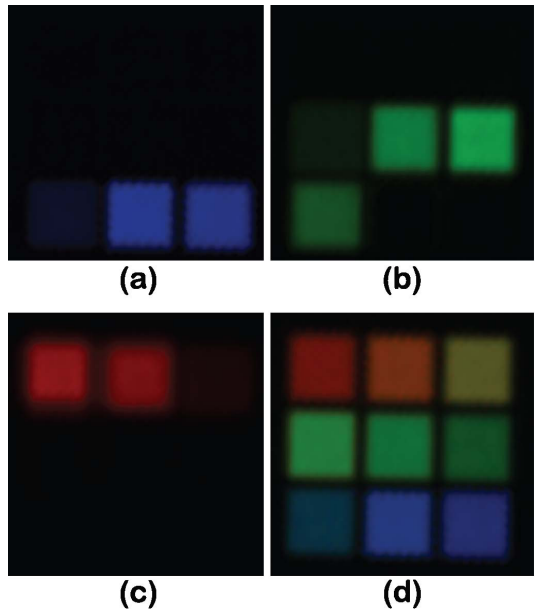}
\caption{Response of a 3\by 3 array of photonic crystal patterns to different inputs. (a) 470nm waveguided light is primarily outcoupled by two array elements in the bottom row, (b) 530nm waveguided light is primarily outcoupled by two array elements in the middle row and one in the bottom row, (c) 630nm waveguided light is primarily outcoupled by two array elements in the top row, and (d) a white light emitting diode (LED) spectrum is outcoupled by array elements in all rows. Each photonic crystal array element measures approximately 30\thinspace \um\by 30\thinspace \um; the total array size is approximately 100\um\by 100\um.}
\end{figure}

\begin{figure}[htbp]
\centering\includegraphics[width=7cm]{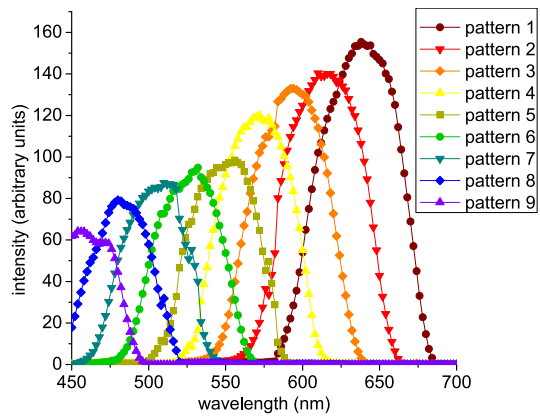}
\caption{Spectral response of photonic crystal patterns. Spectral response of the 3\by 3 array of photonic crystal patterns pictured in Figure 3 measured as mean neutral-weight grayscale intensity vs. wavelength. Peak wavelengths are determined by the photonic crystal lattice constants.}
\end{figure}

The photonic crystal spectrometer works by mapping intensities in pattern space to wavelength space via the response functions for each pattern. Using a matrix formalism, the response of the system can be described as
\begin{equation}
\bm{A} \bm{x} = \bm{b}
\end{equation}
where $\bm{A}$ is an $m$\by $n$ matrix consisting of the intensities of $m$ patterns at $n$ wavelengths, $\bm{x}$ is the wavelength space representation of the input, and $\bm{b}$ is the pattern space representation of the input. Since the $m$ patterns peak at different wavelengths, $\bm{A}$ is full rank therefore its Moore-Penrose pseudoinverse is a valid right inverse \cite{math} which can be used to solve Equation (1) for the spectral response of the input.
The projection operator $\bm{P}$, on to the subset of wavelength space realizable given $\bm{A}$, is
\begin{equation}
\bm{P} = \bm{A^{-1}_{right}} \bm{A}.
\end{equation}
The recovered spectrum $\bm{\tilde{x}} = \bm{P} \bm{x}$ is therefore
\begin{equation}
\bm{\tilde{x}} = \bm{A^{-1}_{right}} \bm{b}.
\end{equation}

Using this formalism, a single photograph containing multiple photonic crystal patterns, such as Figure 3(d), is used to measure the spectral content of a light source. The recovered spectrum in Equation (3) is a linear combination of pattern response functions. The performance of the spectrometer is determined by $\bm{P}$ which is a function of both the number of pattern elements in this basis set and the number of wavelengths used to characterize the patterns. The resolution is constrained by the number of wavelengths used to characterize the patterns, while the accuracy is constrained by the ability of the basis to represent the measured spectrum.

Figure 5 compares the performance of a nine pattern photonic crystal spectrometer to a commercial spectrometer. The major features in the spectrum have been reproduced using a simple and inexpensive system: nine patterns in PMMA on a glass substrate and a camera. The spectrum using the photonic crystal spectrometer is obtained through the analysis of grayscale intensities of the patterns in Figure 3(d), using Equation (3).  The projected spectrum, also shown in Figure 5, represents the performance limit for the photonic crystal spectrometer given the pattern response functions shown in Figure 4; it is obtained by projecting the LED spectrum measured with the commercial spectrometer on to the basis functions shown in Figure 4 using Equation (2). 

\begin{figure}[htbp]
\centering\includegraphics[width=7cm]{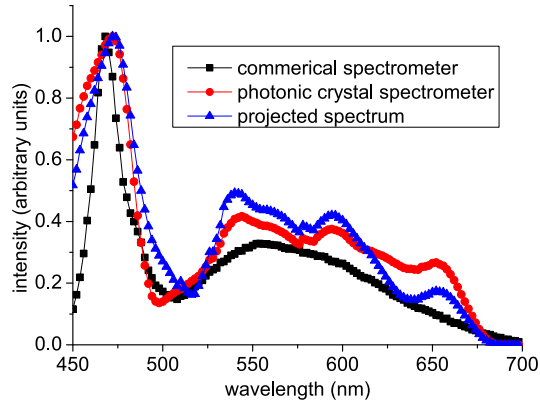}
\caption{Comparison of photonic crystal spectrometer and commercial spectrometer. Comparison of white LED spectrum measurement using commercially available spectrometer, calculated spectrum projected on to the pattern spectral response function basis shown in Figure 4, and measurement using the photonic crystal spectrometer.}
\end{figure}

\begin{figure}[htbp]
\centering\includegraphics[width=7cm]{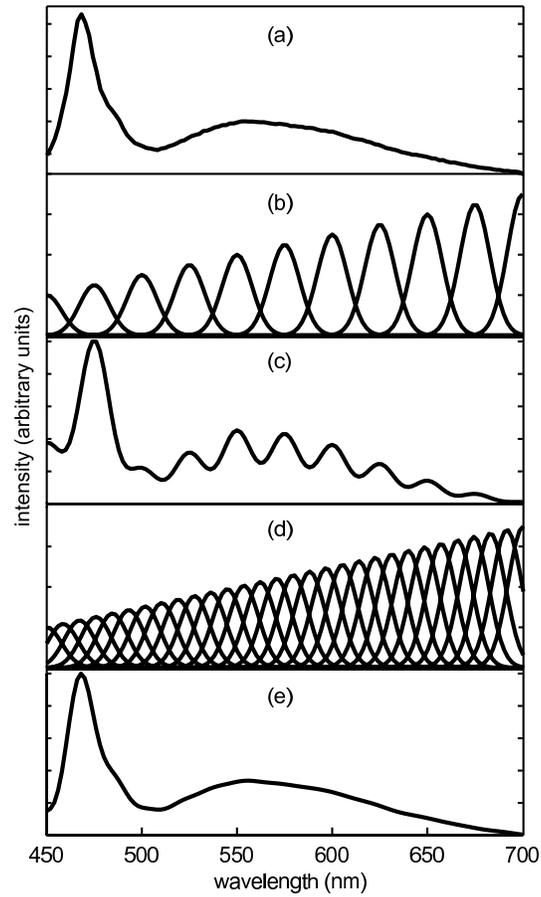}
\caption{Effect of number of basis elements on photonic crystal spectrometer response. Calculated projections of ideal white LED spectrum on to different bases, with 2\thinspace nm wavelength step size. (a) ideal white LED spectrum (no projection), (b) 11 response function basis similar to pattern responses in Figure 4, (c) projection of white LED spectrum on to 11 function basis in (b), (d) 30 response function basis, (e) projection of white LED spectrum on to 30 function basis in (d).}
\end{figure}

The accuracy of the recovered spectrum can be improved with the use of more photonic crystal patterns. Figure 6 shows the effect of an increased density of response functions (similarly shaped to those in Figure 4) on a white LED spectrum projection. The 11 element response function basis in Figure 6(b) is similar to that of the experimental system (Figure 4). Increasing the number of response functions to 30, as in Figure 6(d), greatly improves the accuracy of the projected spectrum. The project spectrum in Figure 6(e) is a faithful reproduction of the unaltered spectrum in Figure 6(a). The experimental realization presented here is an array of nine patterns, but the same measurement setup can handle more than 600 similarly sized patterns without any modification. Thus a higher performance photonic crystal spectrometer is readily attainable.

\section{Conclusion}

We have demonstrated a simple, low cost spectrometer based on the ability of photonic crystals to wavelength selectively outcouple waveguided light from a glass slide. The response of an array photonic crystal patterns on the glass slide to a light source is captured using a camera. The resulting image is used to calculate the spectrum of the light source, via the response functions of the photonic crystal patterns. 

The most expensive component of the photonic crystal spectrometer is the camera. Compact cameras, ubiquitous in cell phones and laptops, are extremely inexpensive; CMOS camera modules retail for less than \$20 \cite{camera-price}. The photonic crystal patterns presented here were fabricated using electron beam lithography, however nanoimprint lithography of patterns with similar features to the highest aspect ratio patterns presented here has previously been demonstrated in the literature with excellent uniformity over 3\thinspace mm\by 3\thinspace mm areas \cite{imprint}. Thus the photonic crystal arrays can be manufactured using an extremely low cost and high throughput technique. With a high quality imprint master, the only limitation on the number of patterns that can be incorporated into the spectrometer is the camera resolution.

One advantage of the photonic crystal spectrometer is that it can be easily tailored to specific applications through the selection of the incorporated photonic crystal patterns. This allows for the use of lower resolution cameras in applications where broadband high resolution performance is not required. For example, in a low cost spectrophotometer designed for the detection of a specific compound, the photonic crystal patterns can be limited to those required for detection of wavelengths of interest and rejection of false positive compounds.

\section{Acknowledgments}
The authors wish to acknowledge support from eMagin under subcontract from the U.S. Army Night Vision \& Electronic Sensors Directorate. This work has used the shared experimental facilities that are supported primarily by the MRSEC Program of the National Science Foundation under Award Number DMR-0213574 and by the New York State Office of Science, Technology and Academic Research (NYSTAR). The authors would like to thank M. S. Foster and L. A. Fetter for helpful discussions.

\end{document}